\documentclass{elbart}
\usepackage{graphicx}

\begin{document}
\begin{frontmatter}
\title{Lorentz angle measurements in irradiated silicon detectors\\
between 77 K and 300 K}
\author[IEKP]{W. de Boer},
\author[IEKP]{V. Bartsch},
\author[IEKP]{J. Bol},
\author[IEKP]{A. Dierlamm},
\author[IEKP]{E. Grigoriev},
\author[IEKP]{F. Hauler},
\author[IEKP]{S. Heising},
\author[IEKP]{O. Herz},
\author[IEKP]{L. Jungermann},
\author[IEKP]{R. Ker\"anen},
\author[IEKP]{M. Koppenh\"ofer},
\author[IEKP]{F. R\"oderer} and
\author[ITP]{T. Schneider}
\address[IEKP]{Institut f\"ur Experimentelle Kernphysik,  Universit\"at
Karlsruhe, Germany}
\address[ITP]{Institut f\"ur Technische Physik, Forschungszentrum
Karlsruhe, Germany}

\begin{abstract}
Future experiments are using silicon detectors in a high radiation
environment
and in high magnetic fields. The radiation tolerance of silicon improves
by cooling it to temperatures below $180\;\rm K$.
At low temperatures the mobility increases, which leads to larger
deflections
of the charge carriers by the Lorentz force.

A good knowledge of the Lorentz angle is needed for design and
operation of silicon detectors. We present measurements of the Lorentz
angle
 between $77\;\rm K$ and $300\;\rm K$ before
and after irradiation with a primary  beam of 21 MeV protons.

\end{abstract}
\end{frontmatter}
\vspace*{\fill}

\section{Introduction}

Future collider experiments need stronger magnetic fields for momentum
measurements, because of the higher particle momenta.
In high fields the drifting charge carriers generated by traversing
particles
are deflected significantly by the Lorentz force
$\rm e\vec{v}\rm{x}\vec{B}$,
where $\rm \vec{v}$ is the drift velocity and $\rm \vec{B}$
the magnetic field.
As advantage of such a deflection one can consider the improved
charge sharing between the readout strips, which can improve
the resolution for a given readout pitch.
On the other hand the charge sharing worsens the double track resolution
and the signal-to-noise ratio.
As is well known\cite{old,roederer:1998,heising:1999,hauler:2000},
the drift mobility of electrons
is larger than the hole mobility, which yields
a considerably  larger Lorentz shift for electrons than for holes.
Therefore, $\rm p^+nn^+$ silicon detector with charge integration on both
sides, show much less of an effect
of the magnetic field on the p-side, where the holes are collected,
than on the n-side, where the electrons are collected.

Typical position resolutions of silicon strip detectors are in the order
of $\mu \rm m$, while the Lorentz shifts in a 4 T magnetic field
reaches 200 $\mu \rm m$ for electrons in a 300 $\mu \rm m$ thick detector.
Therefore, these shifts have to be accounted for and an interesting
question is the dependence of the Lorentz shift on the irradiation dose.
For an LHC experiment, where the dose reaches over $10^{14}$ particles
per $\rm cm^2$, the change in drift mobility due to  the defects
introduced by the radiation damage, might result in a continuous
change of the Lorentz shift, and thus of the calibration of the detector.

The radiation tolerance of silicon detectors
can be improved by either oxygenating the wafers\cite{rd48}
At lower temperatures the mobility of both electrons and holes 
increases rapidly ($\mu_H \propto T^{-2.4}$) and Lorentz angles for
electrons up to 80$^\circ$ were observed in a 4 T magnetic field
at temperatures of 80 K\cite{heising:1999}.

In this paper we study the Lorentz angles of both electrons and holes
in magnetic fields up to 8 T and temperatures between 77 and 300 K.
This is done before and after irradiating a detector with 21 MeV protons
up to a fluence of $\rm 10^{13}/\rm cm^2$, which equals
$\approx 2.8\cdot10^{13}/\rm cm^2$
1 MeV equivalent neutrons.

\section{Experimental setup}

\begin{figure}
\centering{\includegraphics[width=12cm,clip]{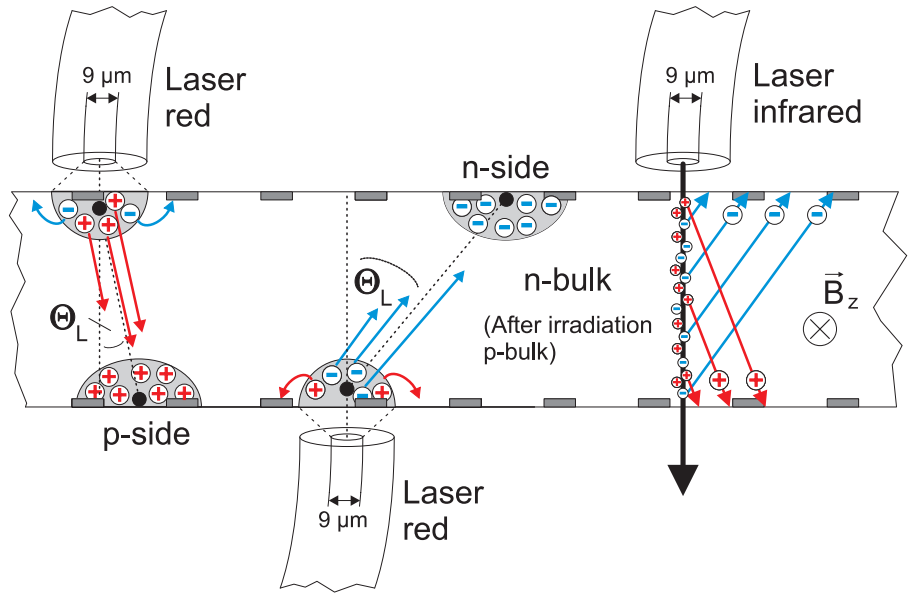}
}
\caption{\label{f1} The detectors have three lasers
connected to them.
The red lasers have a penetration depth of a few $\mu m$, so with
a laser pulse on the n-side or  p-side one can measure the drift
from electrons and holes, respectively. With the infrared laser
one can simulate a through going particle.}
\end{figure}

The Lorentz angle $\Theta_L$ under which charge carriers are deflected
in a magnetic field perpendi\-cular to the electric field is defined by:
\begin{equation}
\tan (\Theta_L) = \frac{\Delta x}{d} = \mu_H B = r_H \mu B
\label{eq1}
\end{equation}
where the drift length corresponds to the detector thickness $d$ and
the shift of the center of charge is $\Delta x$ (see figure \ref{f1}).
The Hall mobility is denoted by $\mu_H$, the conduction mobility
by $\mu$. The Hall mobility differs from the conduction mobility by
the Hall scattering factor $r_H$. This factor describes the influence
of the magnetic field on the  mean scattering time of carriers of
different energy and  velocity\cite{smith}.
The Hall scattering factor has a value of  1.15 (0.7) for electrons (holes)
at room temperature and decreases (increases)  towards $1.0$
with decreasing temperature \cite{lb}.
The mobility $\mu$ increases with temperature proportional
to $T^{-2.42}$ \cite{sze}.

The Lorentz angle can be measured by injecting charges at the surface
on one side and observing the drift through the detector
by measuring the position of this  charge on
the opposite side(see figure \ref{f1}). 
Charges were generated by injecting light with a wavelength
of $\lambda = 650\;\rm nm$, which has an
 absorption length of
$3\; \mu \rm m$ at $300\;\rm K$ and $10\;\rm \mu m$ at $77\;\rm K$.
Alternatively an infrared laser with a wavelength of 1060 nm
was used, which has an absorption length of 300 $\mu \rm m$
at room temperature. This laser penetrates the detector and so mimicks
a minimimum ionizing particle.
With the red laser
one type of carriers immediately recombines at the
nearest electrode, whereas the other type drifts towards the opposite side.
This allows to measure the Lorentz angle for electrons and holes
separately by either injecting laser light on the n- or p-side.

\begin{figure}
\centering{\includegraphics[width=12cm,clip]{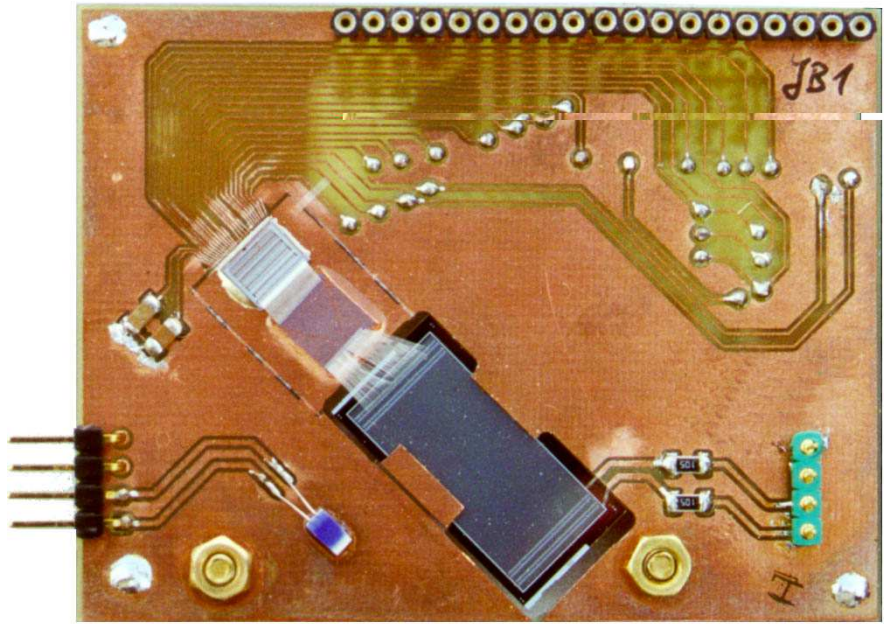}}
\caption{\label{f2}The detector mounted at an angle of 45$^\circ$
on the hybrid in order to be able to measure the Lorenzt shift
on both sides of the detector. The pitch adapter and the Premux128
chip with the 128 charge sensitive amplifiers can be clearly seen.
The strips are parallel to the edges of the detector and oriented
at an angle of $90^\circ$ with respect to each other on  the p- and n-side. 
The magnetic field is directed from
the bottom to the top of the hybrid.}
\end{figure}

For our measurements the JUMBO magnet from the Forschungszentrum
Karlsruhe\cite{jumbo} was used
in a $B =10\;\rm T$ configuration with a warm bore of 7 cm.
A flow of cold nitrogen gas through the warm bore allowed the detectors
to be cooled to temperatures between 77 and 300 K.
The sensors are double sided ``baby'' detectors
of approximately 2x1 cm from the HERA--B
production by Sintef. They have a
 strip pitch of 50 micron on
the p--side and 80 micron on the n--side; the strips on opposite sides
are oriented at an
angle of 90 degree with respect to each other.
The x- and y-directions are taken to be along the strips,
while the E-field is in the z-direction.
The B-field, perpendicular to the electric field, has to be in the x-y
plane,
but cannot be oriented along the x- or y-direction, since then
the Lorentz shift would be along one of the strips, i.e. unmeasurable.
Therefore, the detector is rotated 45 degrees, so that the B-field
direction is at  an angle of 45 degrees with both the x- and y-axis,
as shown in fig. \ref{f2}. The sensor is glued on the hybrid together
with a pitch adapter and 
a 128 channel
Premux charge sensitive amplifiers\cite{jones} on each side.
The conections are standard wirebonds.
All signals are digitized
for every laser pulse
and averaged over a few 100 pulses.

The signal
position is computed using either the center of gravity or  a Gauss-fit:
\begin{equation}
\label{eq:charge-cog} \bar{x} (PH) = \frac{\sum PH_i x_i}{\sum PH_i}.
\end{equation}
Here $PH_i$ is the pulse height of the  strip $i$ and $x_i$ its
position.

\begin{figure}
\centering{
\includegraphics[width=11cm,clip]{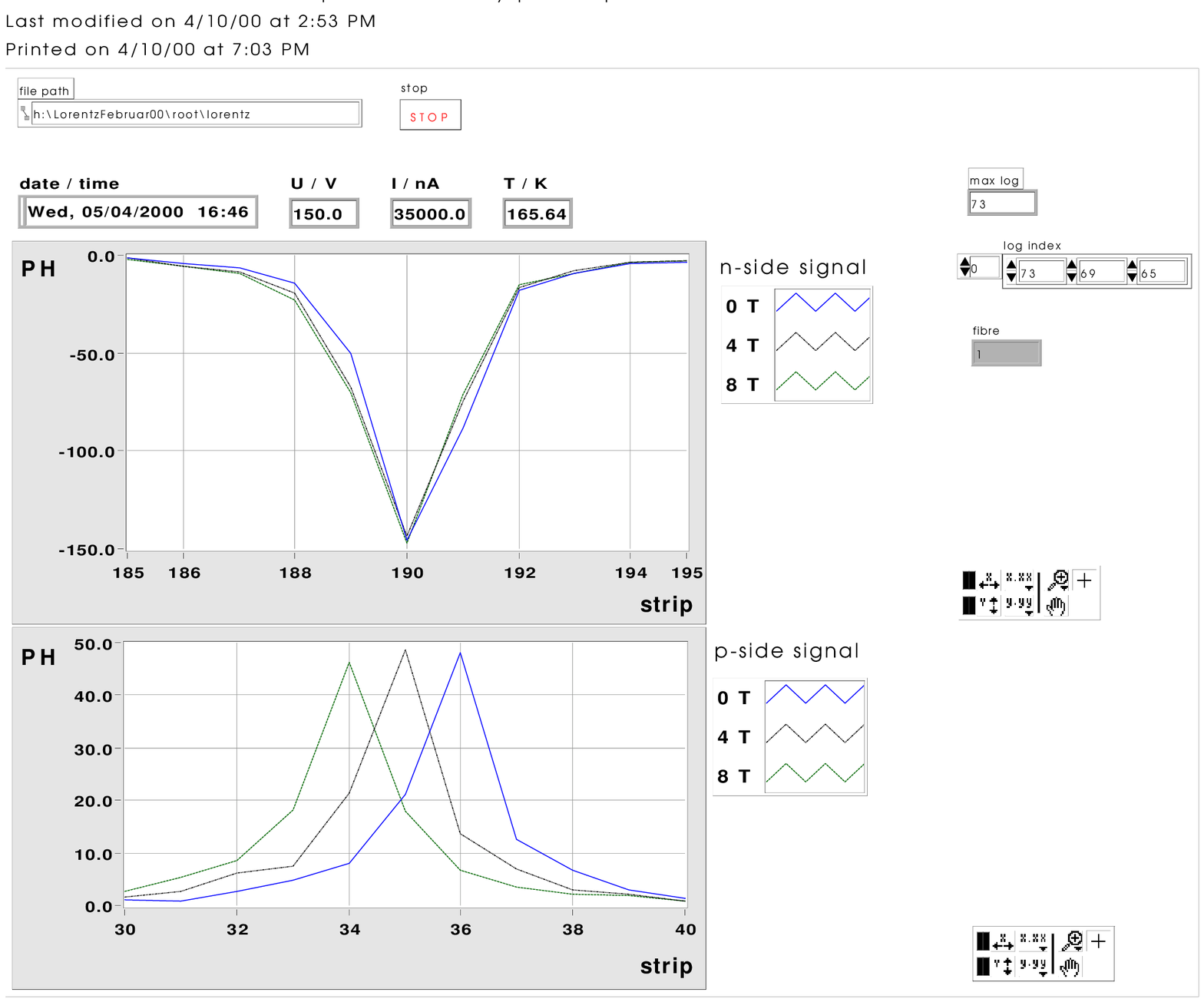}
}\caption{\label{f3}Laser pulse position in different
magnetic fields. Light
is injected from the n-side resulting in an almost stable pulse position
on that side and a clearly moving position on the p-side, if the
magnetic field is increased to 8 T in steps of 4 T.}
\end{figure}

\begin{figure}
\centering{
\includegraphics[width=12cm,clip]{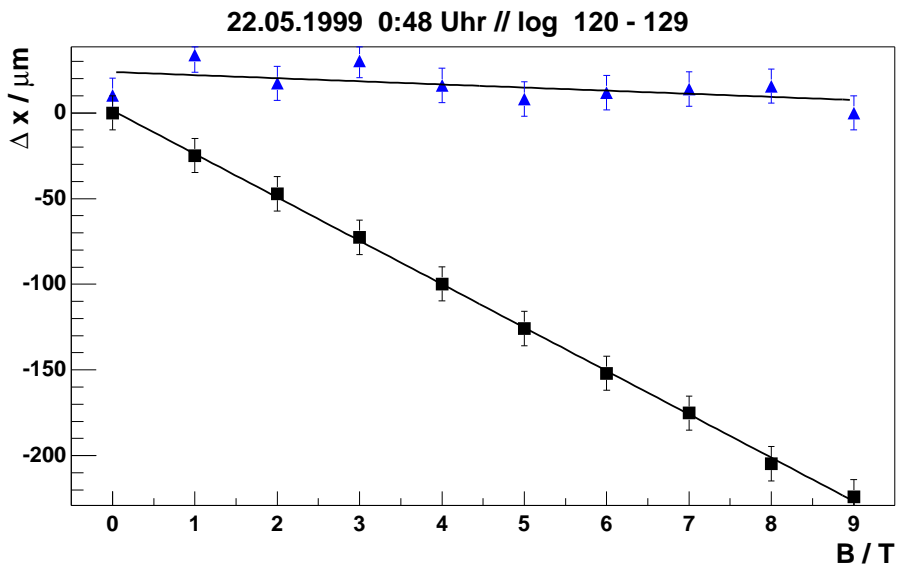}
}
\caption{\label{f4}The Lorentz
shift versus magnetic field. The red laser pulse on the n-side
side hardly penetrates, so the electrons are immediately
absorbed by the neighbouring strips on the n-side and the holes drift
through the detector to  the p-side. Therefore
an appreciable shift is only seen for the holes.}
\end{figure}

As can be seen from
 figure \ref{f3}, the pulse on one side hardly  moves due to
the immediate recombination, while the pulse on the opposite side
 shows a clear Lorentz shift.
The signal position is plotted as a function of the magnetic
field in figure \ref{f4}, which  shows clearly that the Lorentz shift
is linear with the magnetic field up to 9 T.

In order to  better understand the drift in the detector
simulations were performed with 
 the Davinci software package by TMA\cite{davinci}.
The inhomogeneous electric field in the sensor was taken into account
by following the charge in small steps, calculating the mobility
at each position and integrating the Lorentz shift using eq. \ref{eq1}
with a Hall scattering factor of $1.15$ and $0.7$ for
electrons and holes, respectively. These are the values
expected for highly  resistive sensors, in which the scattering
by phonons dominates over the scattering by impurities\cite{sze}.

\begin{figure}
{\centering
\includegraphics[clip,width=0.45\textwidth]{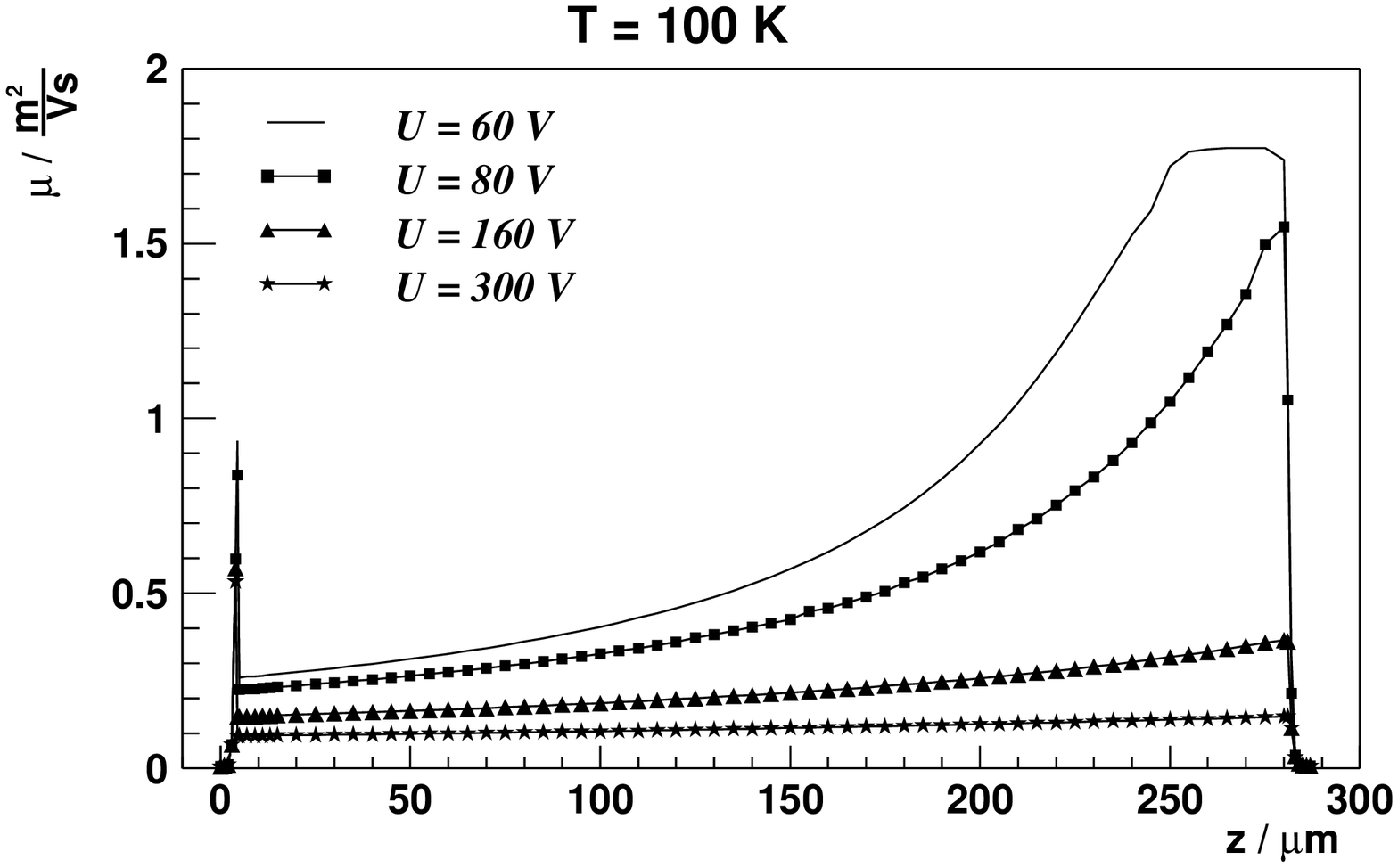}
\hfill
\includegraphics[clip,width=0.45\textwidth]{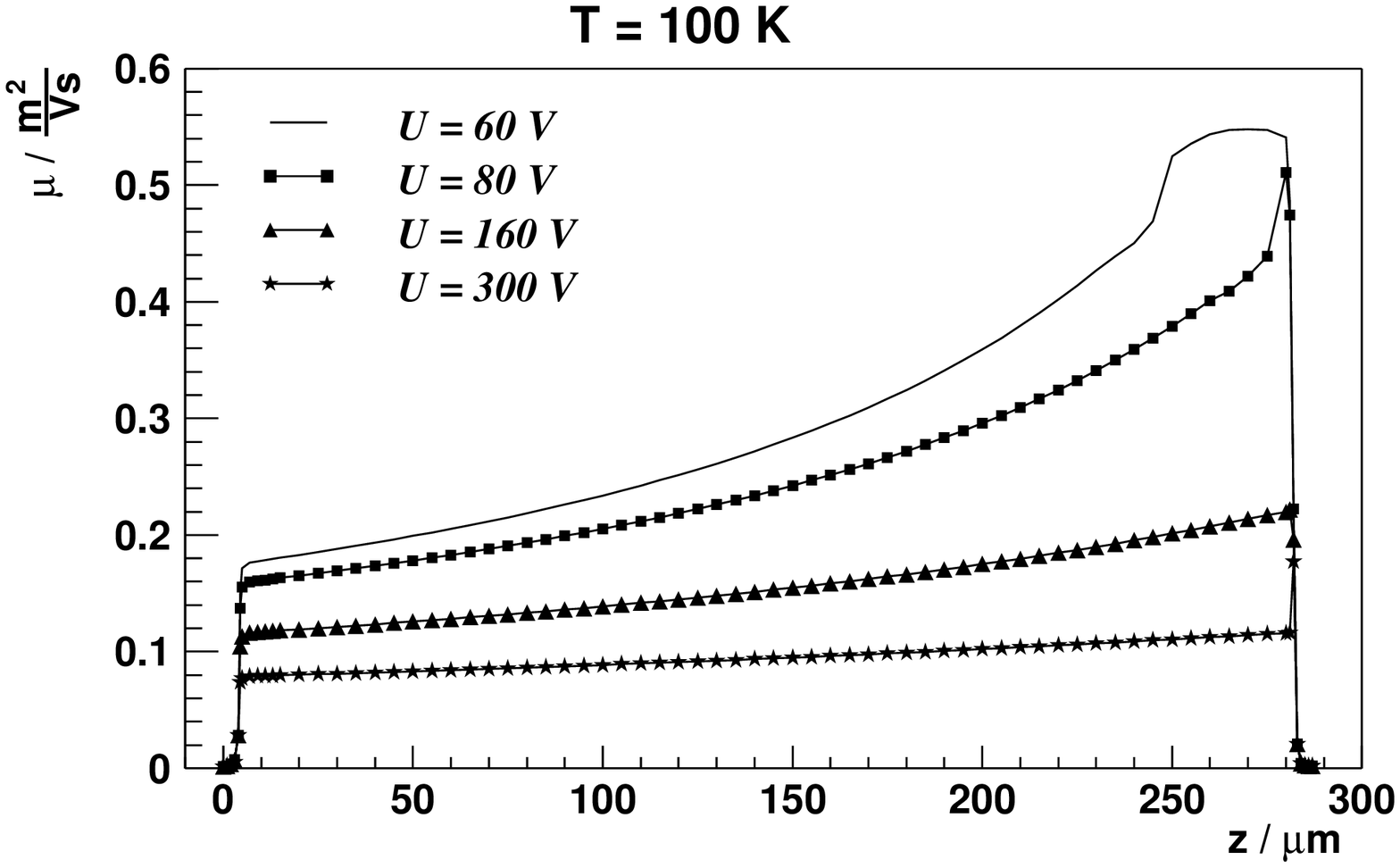}
}
{\centering
\includegraphics[clip,width=0.45\textwidth]{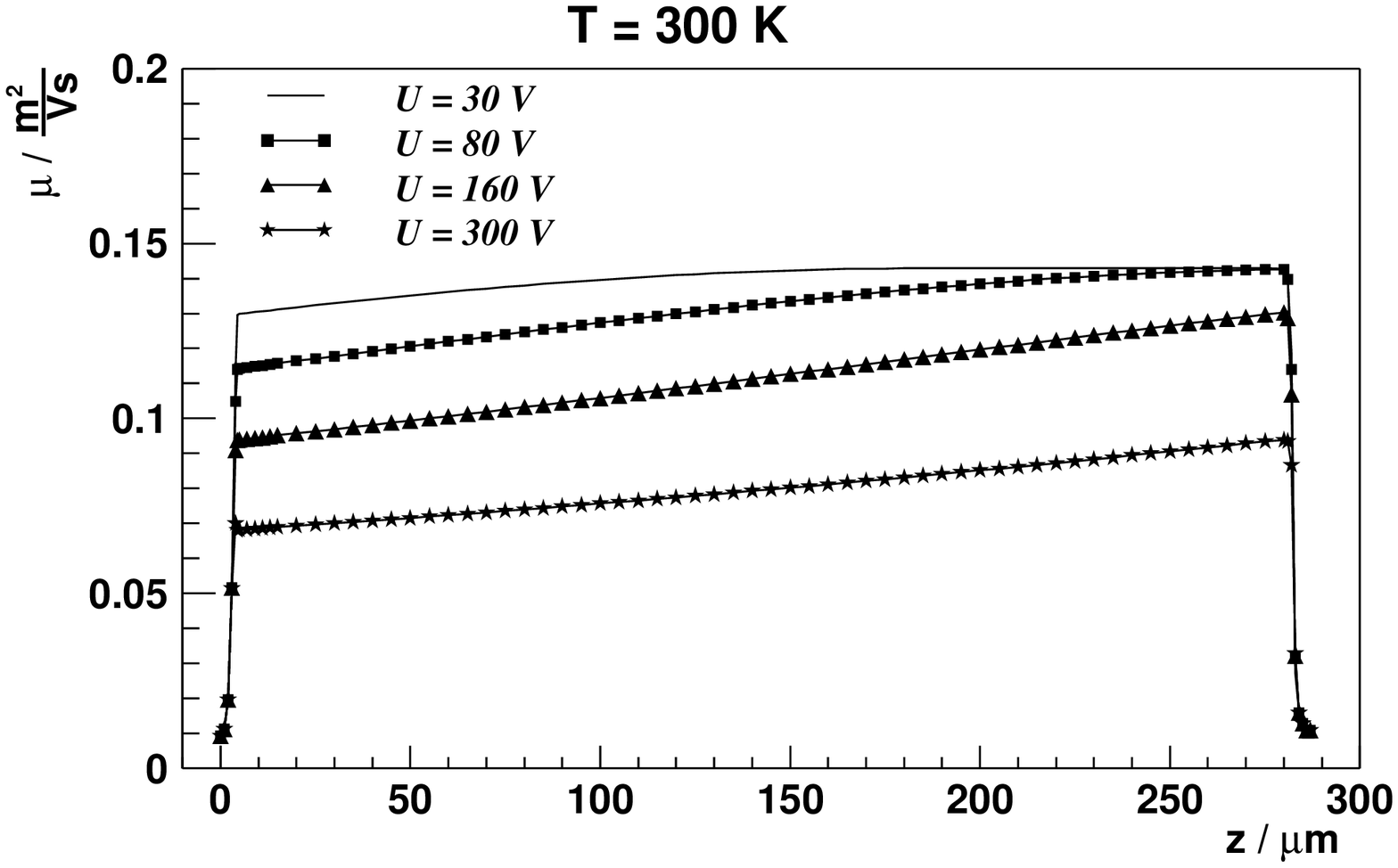}
\hfill
\includegraphics[clip,width=0.45\textwidth]{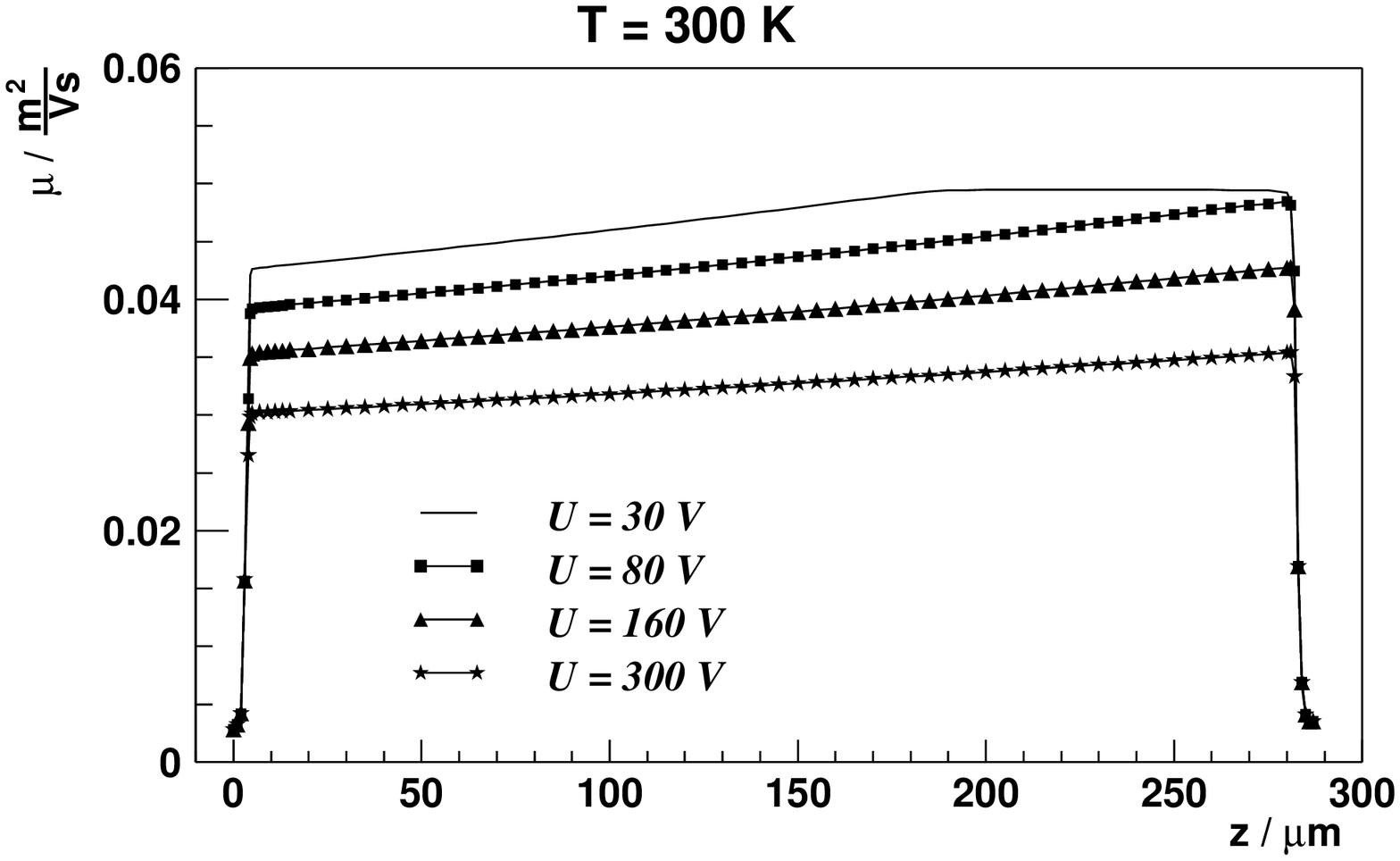}
}
\caption{\label{f5} Mobilities of electrons (left) and holes (right)
 for a detector with a full depletion voltage of 80 V.
The top row is for a temperature of 100 K, the lower row for 300 K.
From \cite{heising:1999}.}
\end{figure}

\begin{figure}
{\centering
\includegraphics[clip,width=0.7\textwidth]{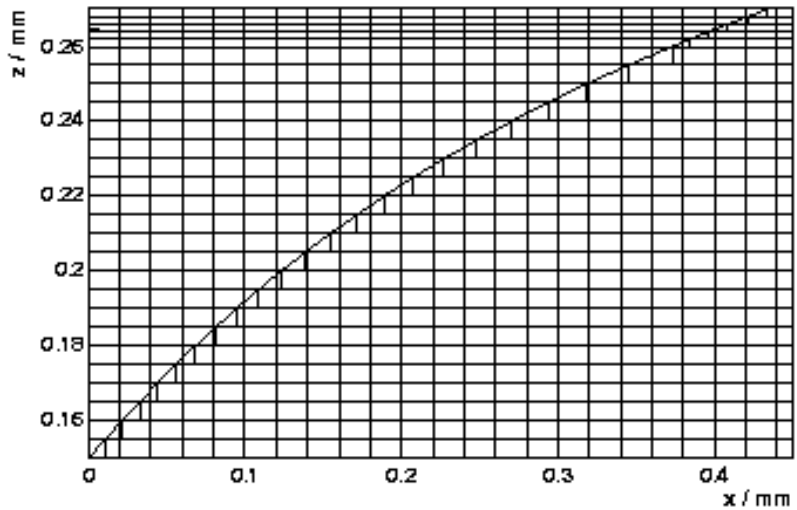}\\}
\caption{\label{f6} Mean trajectory of charge in a just depleted
detector at a temperature of 80 K in a 4 T magnetic field.
From \cite{heising:1999}.}
\end{figure}

For bias voltages at least a factor two  above the depletion voltage the 
mobility is practically constant in the detector, while
for just depleted detectors the mobility $\mu=v/E$  increases in the
regions, where the electric field goes to zero, especially at low temperatures,
as shown in figure \ref{f5}.
The corresponding mean trajectory of the ionization becomes
non linear, as shown in figure \ref{f6}.

\section{Results}

\begin{figure}
\centering{
\includegraphics[width=14cm,clip]{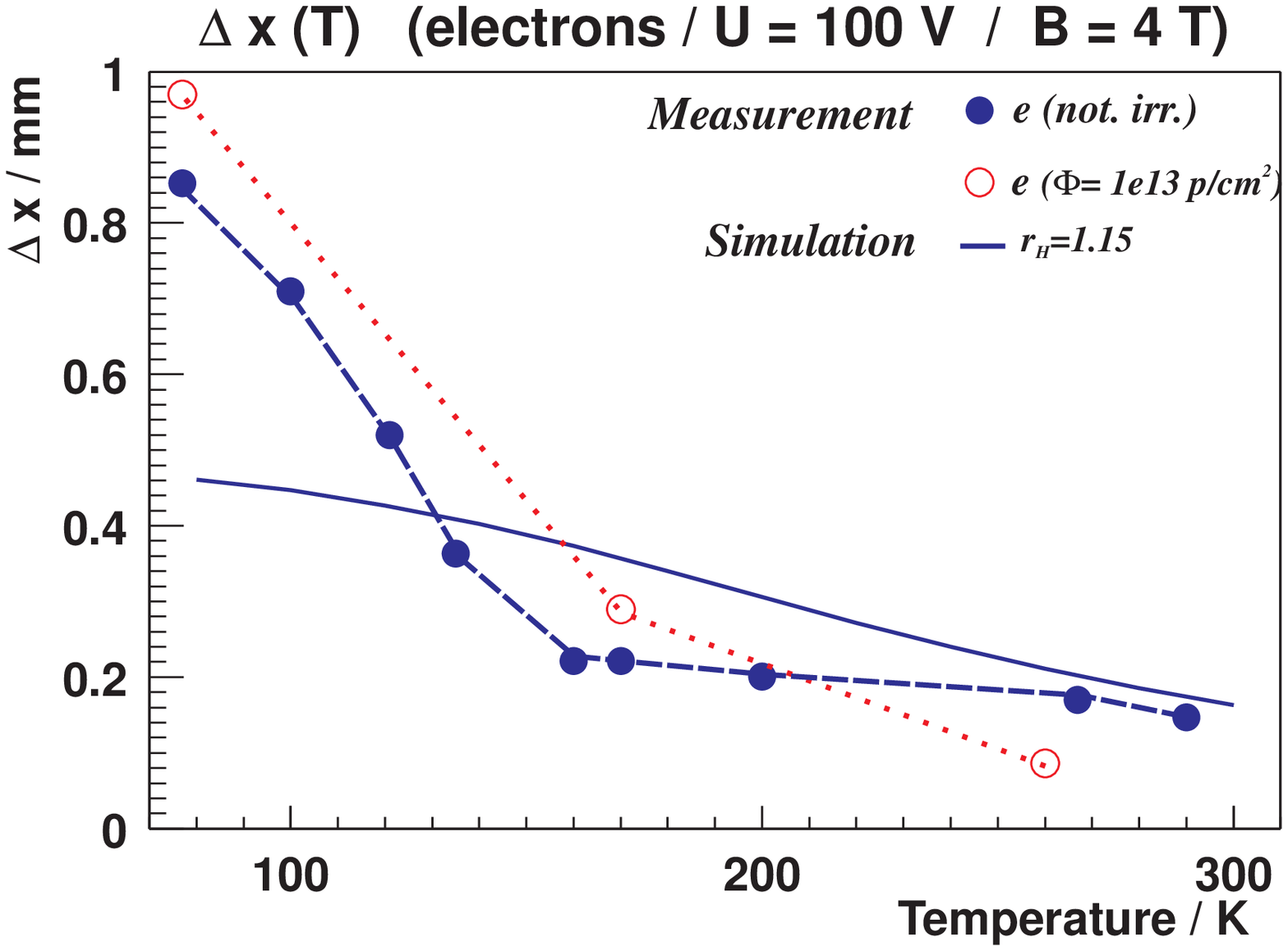}
\includegraphics[width=14cm,clip]{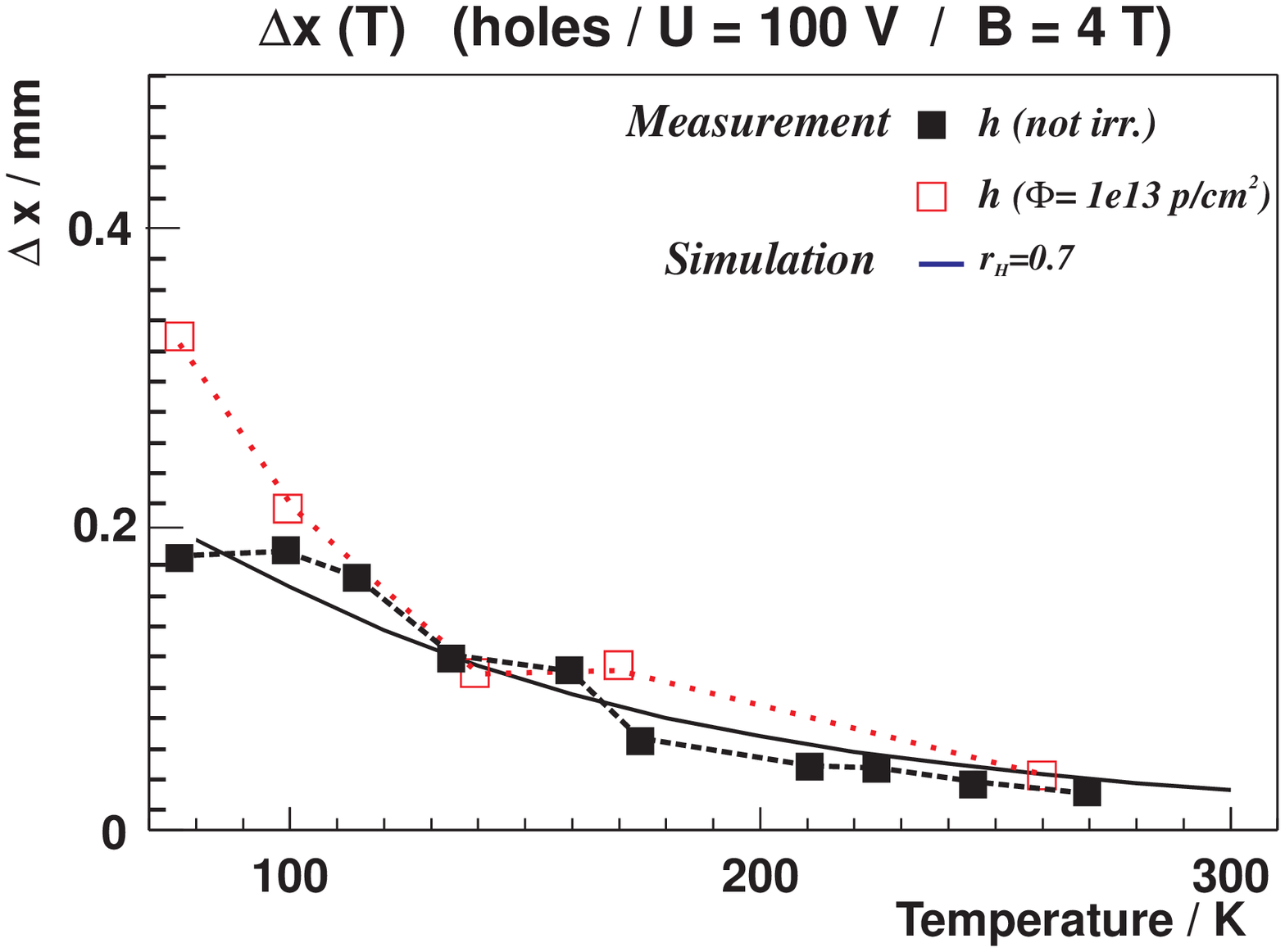}
}
\caption{\label{f7}Lorentz shift for electrons (top)
and holes (bottom) for 300 $\mu m$ detector
 in a 4 T magnetic field as a function of
temperature. Data  for an
unirradiated detector  and a detector irradiated with protons
of $21\;\rm MeV$ to a fluence of $1.0\cdot 10^{13}~{p}/{cm^2}$ are shown.
For comparison, the temperature dependence from the DaVinci simulation
program with a constant Hall scattering factor of 1.15 (0.7)
for electrons (holes) is shown
by the full lines.}
\end{figure}

\begin{figure}
\centering{
\includegraphics[width=14cm,clip]{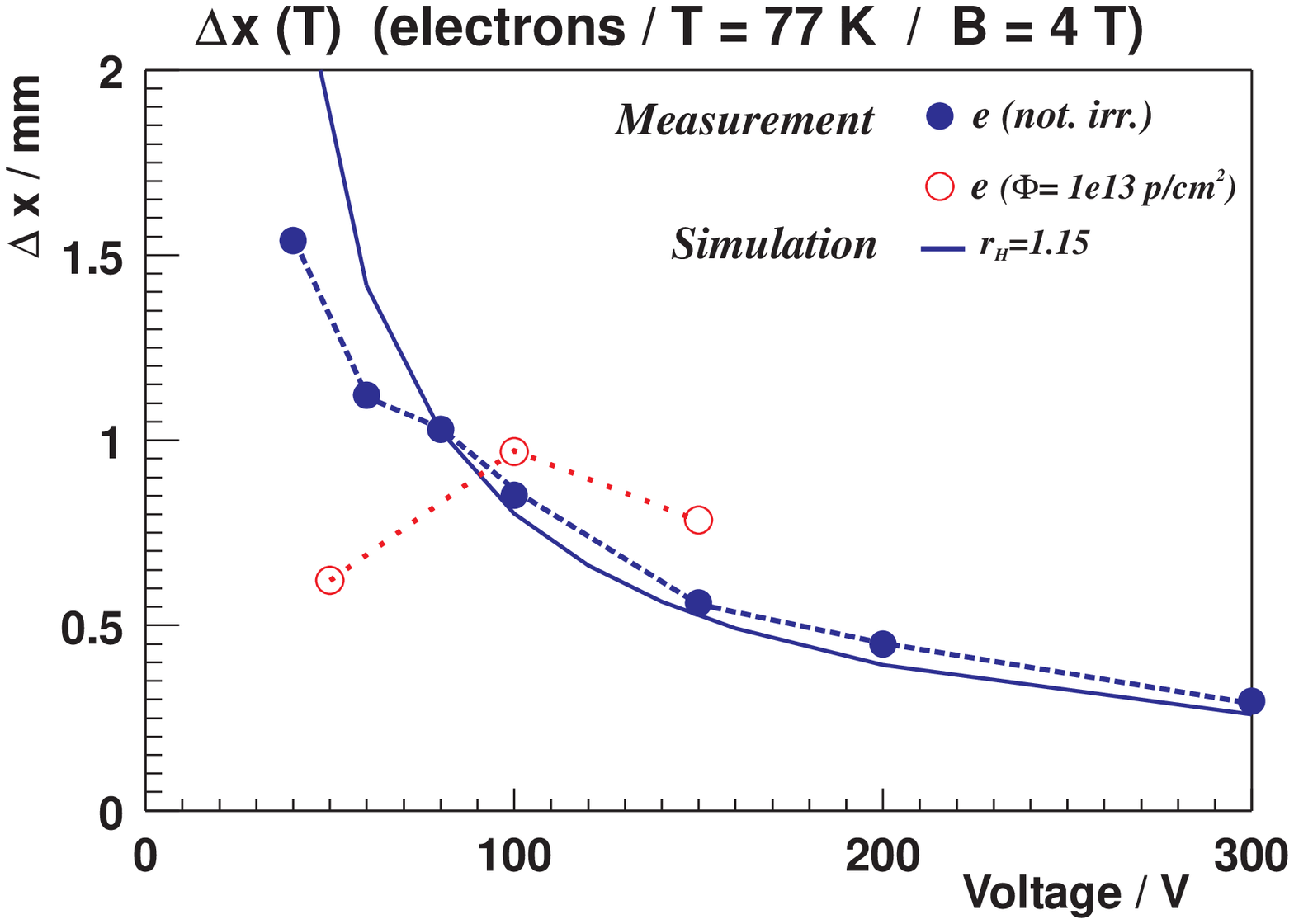}
\includegraphics[width=14cm,clip]{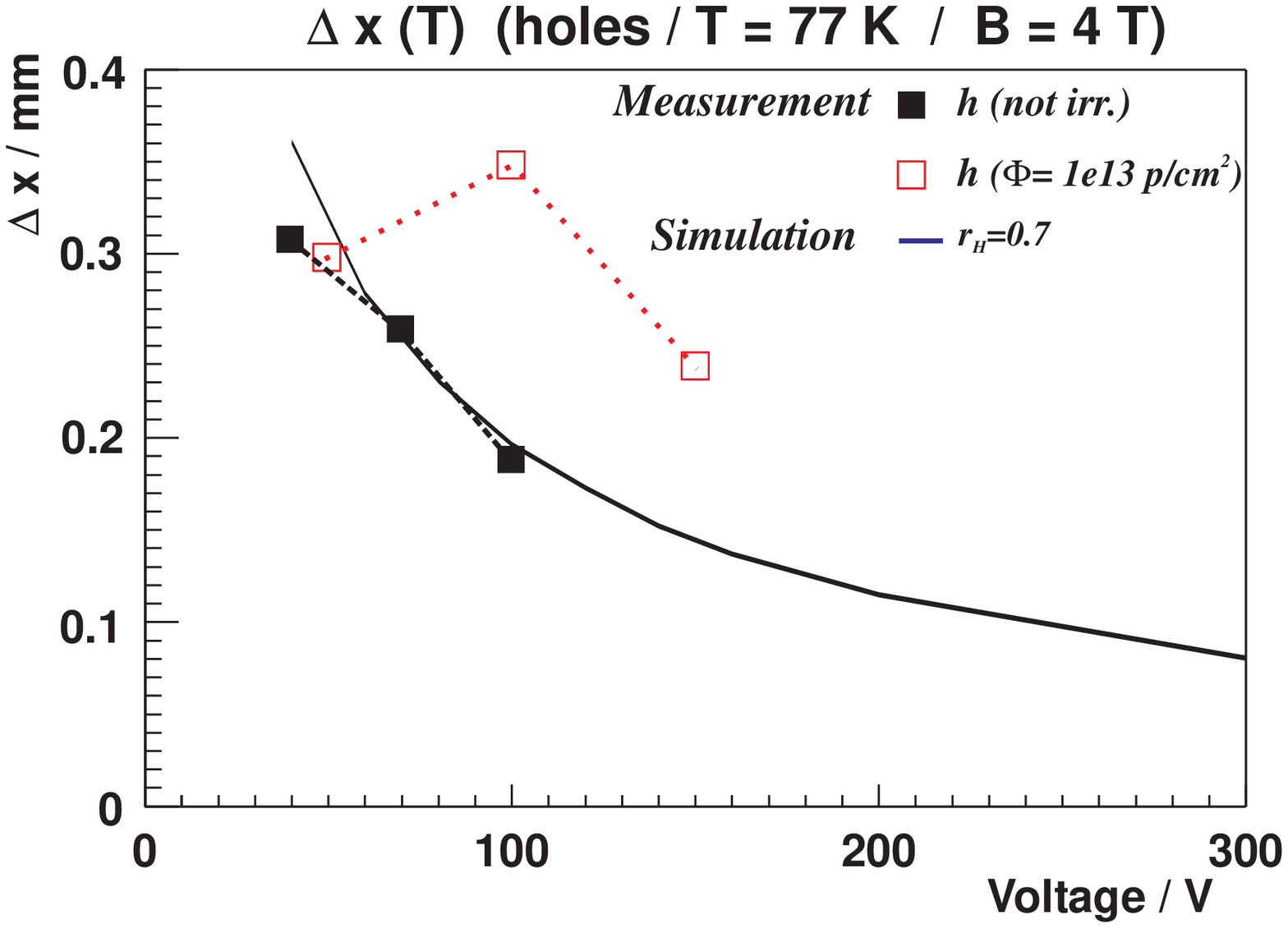}
}
\caption{\label{f8}Lorentz shift for a 300 $\mu m$ detector
 in a 4 T magnetic field
 for electrons (top) and holes (bottom)  versus bias voltage at a
temperature of 77 K. Both, data for an
unirradiated detector  and a detector irradiated with protons
of $21\;\rm MeV$ to a fluence of $1.0\cdot 10^{13}~ {p}/{cm^2}$ are shown.
For the latter the full depletion voltage has increased from 40 to 100V.
For comparison, the temperature dependence from the DaVinci simulation
program with a constant Hall scattering factor of 1.15 (0.7)
for electrons (holes)is shown by the full
lines.}
\end{figure}

Instead of the Lorentz angle the shift in a $300 \mu\rm m$
thick detector is plotted for a 4 T magnetic field, which
is the one of interest for future experiments.
The dependence on temperature and bias voltage is shown in
figures \ref{f7} and
\ref{f8}  together with the simulations from
 the Davinci software package by TMA\cite{davinci}, as mentioned above.

For holes the temperature dependence is well described,
but for  electrons the Lorentz angle first falls below the simulation
for decreasing temperature,
as expected for a decreasing value of the Hall scattering factor $r_H$
at lower temperatures\cite{lb}.
However, below $T=160 $ K the Lorentz angle for electrons rapidly
increases and is a factor two $\it above$ the simulation
at liquid nitrogen temperature.
The simulation was done with  a temperature independent
Hall factor of 1.15 (0.7) for electrons (holes), which is certainly
wrong\cite{lb}.
Therefore,
the most likely interpretation of the deviation between simulation
and data would be  the temperature dependence
of the Hall scattering factor, since the drift mobility at
low temperatures is well known\cite{lb} and the electric field
dependence is well described by the simulation, as shown in figure \ref{f8}.
However, if the data is a factor two above the simulation at low temperature,
then a Hall
scattering factor around two is needed.
Such a large $r_H$ is expected for impurity scattering\cite{smith},
but this cannot be reconciled with the high resistivity silicon
used for the sensors, for which the impurity doping is less then
$10^{12}/\rm cm^3$.

Before irradiation  the detector depletes fully with a bias voltage
of 40 V, while after radiation with $1.0\cdot 10^{13}$
21 MeV protons per $\rm cm^2$
the depletion voltage has increased to 100 V. This implies that
the bulk is inverted from n-type to p-type material, as expected\cite{rd48}.
The bulk damage of 21 MeV protons is  about 2.8 times the
damage by 1 MeV neutrons.
The decrease of the Lorentz shift for the irradiated sample below 100 V
in figure \ref{f8} is due to the reduced effective thickness of the partially
depleted detector.
Numerical results on the Lorentz angles and Lorentz shifts have been
summarized in  tables
\ref{tab1} and \ref{tab2} before and after irradiation, respectively.

\begin{figure}
{\centering 
\includegraphics[clip,width=0.85\textwidth]{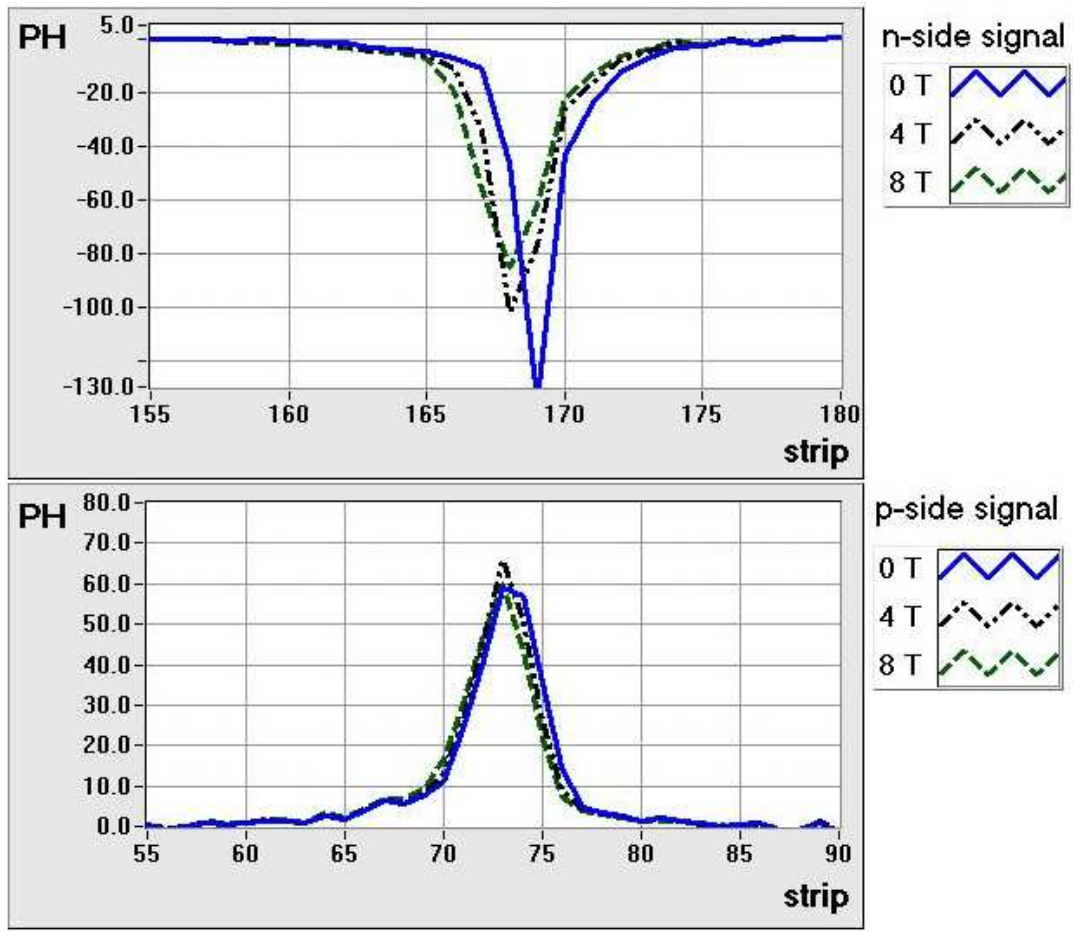}}
\caption{\label{f9} Pulse shape of the infrared laser 
 shining through the irradiated detector ($10^{13}~p/\rm cm^2$).
The n-side signal is mainly
dominated by electrons, the p-side signal by holes, as is apparent
from the displacement as function of the magnetic field.
 }
\end{figure}

Figure \ref{f9} shows the signal from the 1060 nm infrared laser on the
irradiated detector. As mentioned before, this laser has an
absorption length of about 300 $\mu \rm m$ at room temperature,
so it traverses  the detector. It can be seen that the signal
on the p-side hardly moves with the magnetic field, as expected for
a dominant contribution of holes by the charge integrating amplifier,
while the n-side shows a Lorentz shift corresponding to roughly
half the Lorentz shift of the red laser. This is expected for
ionization distributed in the detector, so the average drift length
is half of the detector thickness.

\section{Conclusion}

The Lorentz angle has been  measured for electrons and holes separately.
For the non-irradiated detector the Lorentz angles of holes
agree with simulations at all temperatures, but for electrons
the Lorentz angle agrees only at room temperature.
Below $\approx 160\;\rm K$ the angle increases rapidly to
twice the expected value. A possible explanation could be a Hall
scattering factor of $\approx 2$ instead of $\approx 1$,
although this is in contrast to the expectation for our high
purity silicon sensors, where one expects $r_H=1$ to approach one
at low temperatures.

After irradiation, the Lorentz angle for holes  is hardly changed
at room temperatures, at least if the bias voltage is raised to obtain
full depletion again.
However, for electrons the angle decrease 25\% at room temperature.
At lower temperatures radiation damage increase the Lorentz angles,
both of electrons and holes.

Silicon detectors in high magnetic fields at cryogenic
temperatures can be used without problems, if only the p-side (holes)
is read-out. For
an n--side read-out the too large Lorentz angles can be avoided,
if the strips are oriented such, that the Lorentz shift is parallel
to the strips. If this is not wanted, the Lorentz angle can be reduced
by overbiasing the detector.

\section{Acknowlegdements}

This work was done within the framework of the  RD39
Collaboration\cite{rd39}.
We thank Dr. Iris Abt from the MPI, Munich, Germany for supplying us
with double sided strip detectors from the HERA-B production by Sintef.

\begin{table}[ht]
\begin{center}
\begin{tabular}{|c|cc|cc|cc|cc|}
\hline
& \multicolumn{2}{|c|}{Electrons~(280 K)}
&\multicolumn{2}{|c|}{Holes~(270 K)}
& \multicolumn{2}{|c|}{Electrons~(77 K)} &\multicolumn{2}{|c|}{Holes~(77
K)}\\
Bias / V & $\Theta$ &$\Delta x[\mu m]$  & $\Theta$ &$\Delta x~[\mu m]$
& $\Theta$ &$\Delta x[\mu m]$  &  $\Theta$ &$\Delta x~[\mu m]$ \\
\hline
40 & $33^\circ$     &192&  $6.5^\circ$&34&$79^\circ$&1539&$46^\circ$
&309  \\
100 & $30^\circ$    &170&  $7.2^\circ$&38&$71^\circ$&852 &$32^\circ$ &184
\\
200 &               &   &  $5.9^\circ$&31&$56^\circ$&449 &     &  \\
300 &               &   &  $4.8^\circ$&25&$45^\circ$&295 &     &  \\
\hline
\end{tabular}\vspace*{2mm}
\caption[]{\label{tab1}The Lorentz angle $ (\Theta_L)$
and displacement $\Delta x$
for a 300 $\mu m$ thick detector in a 4 T magnetic field
as function of  bias voltage  at room temperature  and
nitrogen temperature.}
\vspace*{5mm}
\begin{tabular}{|c|cc|cc|cc|cc|}
\hline
& \multicolumn{2}{|c|}{Electrons~(280 K)}
&\multicolumn{2}{|c|}{Holes~(260 K)}
& \multicolumn{2}{|c|}{Electrons~(77 K)} &\multicolumn{2}{|c|}{Holes~(77
K)}\\
Bias / V & $\Theta$ &$\Delta x[\mu m]$  & $\Theta$ &$\Delta x~[\mu m]$
& $\Theta$ &$\Delta x[\mu m]$  &  $\Theta$ &$\Delta x~[\mu m]$ \\
\hline
50 & $21^\circ$     &117&  $8.5^\circ$&45&$65^\circ$&630&$45^\circ$ &297 
\\
100 & $23^\circ$    &127&  $8.0^\circ$&42&$73^\circ$&970&$48^\circ$ &329
\\
150 & $23^\circ$    &126&  $7.6^\circ$&40&$69^\circ$&785&$37^\circ$ &228 
\\
\hline
\end{tabular}\vspace*{2mm}
\caption[]{\label{tab2}As in table \ref{tab1}, but now for a
detector irradiated with 21 MeV protons up to a fluence of
$10^{13}/\rm cm^2$.
}
\end{center}
\end{table}

\newpage

%
\end{document}